\documentclass[10pt]{article}
\usepackage[french,english]{babel}
\usepackage{amsmath}
\usepackage{amsfonts}
\usepackage{amssymb}
\usepackage{times}
\begin{document}
%
%

November 29th, 2012 \hfill
\vskip 4.0cm
{\baselineskip 12pt
\begin{center}
{\bf UNLOCKING THE STANDARD MODEL}
\end{center}
\begin{center}
{\bf II .\quad  1\ \ GENERATION OF QUARKS . \ MASSES AND COUPLINGS}
\end{center}
}
\baselineskip 16pt
\vskip .2cm
\centerline{B.~Machet
     \footnote[1]{LPTHE tour 13-14, 4\raise 3pt \hbox{\tiny \`eme} \'etage,
          UPMC Univ Paris 06, BP 126, 4 place Jussieu,
          F-75252 Paris Cedex 05 (France),\\
         Unit\'e Mixte de Recherche UMR 7589 (CNRS / UPMC Univ Paris 06)}
    \footnote[2]{machet@lpthe.jussieu.fr}
     }
\vskip 1cm

{\bf Abstract:} We continue investigating the Standard Model for one
generation of fermions and two parity-transformed Higgs doublets $K$ and
$H$ advocated for in a previous work \cite{Machet1},
using the one-to-one correspondence, demonstrated there, between their
components and bilinear quark operators. We
show that all masses and couplings, in particular those of the two
Higgs bosons $\varsigma$ and $\xi$, are determined by low energy considerations.
The mass of the ``quasi-standard'' Higgs boson, $\xi$, is
$m_\xi \approx  \sqrt{2}\,m_\pi$; it is coupled to $u$ and $d$ quarks
with identical strengths.  The mass of the lightest one, $\varsigma$, is
$m_\varsigma \approx
m_\pi \displaystyle\frac{f_\pi}{\sqrt{2}m_W/g} \approx\ 68\,KeV$;
it is very weakly coupled to matter except  hadronic matter. 
The ratio of the two Higgs masses is that of the two scales involved in the
problem, the weak scale  $\sigma=\frac{2m_W}{g}\approx 250\,GeV$ and the  chiral
scale $v=f_\pi$,
which  are also, up to a factor $1/\sqrt{2}$,
 the respective vacuum expectation values of the two
Higgs bosons. They can freely coexist and be accounted for.
The dependence of $m_\varsigma$ and $m_\xi$ on $m_\pi$, that is, on quark masses, 
suggests their evolution when more generations are added.
Fermions get their masses from both Higgs multiplets.
The theory definitely stays in the perturbative regime.

\smallskip


\section{Introduction}
\label{section:intro}
The extension of the Standard Model \cite{GSW} for one generation of fermions
advocated for in
\cite{Machet1} is endowed with two Higgs doublets, a ``chiral'' doublet
\begin{equation}
K = \left(\begin{array}{c}{\mathfrak p}^1-i{\mathfrak p}^2 \cr -({\mathfrak
s}^0+{\mathfrak p}^3)
\end{array}\right), \quad <{\mathfrak s}^0>=\frac{v}{\sqrt{2}},
\label{eq:K}
\end{equation}
and a ``weak'' doublet
\begin{equation}
H = \left(\begin{array}{c}{\mathfrak s}^1-i{\mathfrak s}^2 \cr -({\mathfrak
p}^0+{\mathfrak s}^3)
\end{array}\right), \quad <{\mathfrak s}^3>=\frac{\sigma}{\sqrt{2}},
\label{eq:H}
\end{equation}
both isomorphic to the Higgs doublet of the Standard Model
\cite{GSW}. It constitutes the ``smallest maximal extension'' of the
Glashow-Salam-Weinberg model. It is maximal in the sense that it incorporates all
possible $J=0$ scalars (and pseudoscalars) that are expected for a given
number of generations, and it is the smallest extension because it does not
invoke {\em a priori} any physics ``beyond the Standard Model'' now any new
type of particle.

${\mathfrak s}^0$ and ${\mathfrak s}^3$ have non-vanishing vacuum
expectation values (VEV's) as written in (\ref{eq:H}). In there,
the symbols ``$\mathfrak s$'' and ``$\mathfrak p$'' stand respectively for
``scalar'' and ``pseudoscalar'', such that $H$ and $K$ are parity
transformed of each other. Their components that we call generically $h^0,
h^1, h^2, h^3$ transform respectively by $SU(2)_L$ and $SU(2)_R$ according to
\begin{equation}
\begin{array}{rcl}
T^i_L\,.\,h^j&=&-\frac{1}{2}\left(i\,\epsilon_{ijk}h^k +
\delta_{ij}\,h^0\right),\cr
T^i_L\,.\,h^0 &=& -\frac{1}{2}\, h^i,
\end{array}
\label{eq:ruleL}
\end{equation}
and
\begin{equation}
\begin{array}{rcl}
T^i_R\,.\,h^j&=&-\frac{1}{2}\left(i\,\epsilon_{ijk}h^k -
\delta_{ij}\,h^0\right),\cr
T^i_R\,.\,h^0 &=& +\frac{1}{2}\, h^i.\end{array}
\label{eq:ruleR}
\end{equation}

The main steps of this works are the following.
In section \ref{section:wmass} we give the general formula for the mass
of the $\vec W$ gauge bosons in terms of the two VEV's
$<{\mathfrak s}^0>$ and $<{\mathfrak s}^3>$.
In section \ref{section:yukawa} we introduce Yukawa couplings of quarks to
both Higgs doublets $K$ and $H$. It could have
looked more natural to first introduce the scalar potential, but it turns
out that the latter gets strongly constrained by the former. After giving
their general expression, from which we extract the $u$ and $d$ quark
masses in terms of $<{\mathfrak s}^0>$ and $<{\mathfrak s}^3>$, we
investigate in section \ref{section:lowen}
 their low energy limit by using the one-to-one correspondence demonstrated
in \cite{Machet1} between $K$, $H$ and 4-sets of bilinear quark operators.
At this limit, renormalizability is not a concern  and Yukawa
couplings can be rewritten in a very simple form in which, in particular,
symmetries  clearly show up. Using the
Partially Conserved Axial Current hypothesis (PCAC)
\cite{Dashen} \cite{Lee} \cite{dAFFR} and the
Gell-Mann-Oakes-Renner (GMOR) \cite{GMOR} \cite{dAFFR} relation enables to account
for the mass of the pions and to determine the values of all but one
Yukawa parameters. The last one is obtained by
identifying the Goldstones of the spontaneously broken weak  $SU(2)_L$
symmetry. A last constraint results from considering the $\pi^0-\eta$
system and requesting that it be devoid of any tachyonic state. This determines the
quantity $(m_u-m_d)<\bar u u -\bar d d>$ ($(m_u+m_d)<\bar u
u + \bar d d>)$ is determined by the GMOR relation).
We comment at length on fermion masses, and the important role of both
Higgs doublets in their generation.
After gathering the values of the parameters in section
\ref{section:paramsummary},
section \ref{section:pot} is devoted to the scalar potential. $V(K,H)$ is chosen
to be invariant by the chiral group $U(2)_L \times U(2)_R$, which clearly
identifies the Goldstone of chiral symmetry breaking. It only depends on
two parameters, one quadratic and one quartic coupling. At low energy, it
receives corrections from the bosonised (low energy) form of the Yukawa
couplings, which yields an effective potential $V_{eff}(K,H)$. A last
constraint comes from minimizing $V_{eff}$ at the known VEV's of the two
Higgs bosons, which reproduce the pion and $W$ masses. It determines the
value of the quartic coupling and the masses of the two Higgs bosons.
In section \ref{section:couplings},
 we determine their couplings to quarks, gauge bosons and leptons.
Section \ref{section:symmetries} provides some additional considerations concerning
symmetries, Goldstone and pseudo-Goldstone bosons. Several symmetries are at work
and some fields play dual roles. We focus in particular on the custodial
$SU(2)$ symmetry and on the respective roles of $<\bar u u + \bar d d>$ and
$<\bar u u-\bar d d>$. Section \ref{section:moregen} gives some remarks concerning
more generations. Section \ref{section:conclusion} is a brief conclusion.

\section{Kinetic terms for the Higgs doublets and gauge boson masses}
\label{section:wmass}
The masses of gauge bosons arise from the kinetic terms
\begin{equation}
\Big( (D_\mu K)^\dagger D^\mu K + (D_\mu H)^\dagger D^\mu H \Big)
\label{eq:kinetic}
\end{equation}
for the two Higgs doublets $K$ and $H$.
$D_\mu$ is the covariant derivative with respect to the group $SU(2)_L$ of
weak interactions. Owing to the laws of transformations (\ref{eq:ruleL}),
the VEV's of ${\mathfrak s}^0$ and ${\mathfrak
s}^3$ generate a mass $m_W$ for the $\vec W$ gauge bosons
\begin{equation}
m_W^2 =\frac{g^2}{2}\big(<{\mathfrak s}^0>^2 + <{\mathfrak s}^3>^2\big)
= g^2\, \frac{v^2 + \sigma^2}{4},
\label{eq:mw}
\end{equation}
in which $g$ is the $SU(2)_L$ coupling constant
\begin{equation}
g
\approx .61\ .
\label{eq:g}
\end{equation}

\section{Yukawa couplings}
\label{section:yukawa}

We choose to first introduce Yukawa couplings because their low-energy limit
(see section \ref{section:lowen})
will in particular constrain the effective scalar potential.

\subsection{General expression}
\label{subsec:genyuk}

Quarks must be coupled to the two Higgs doublets $K$ and $H$. Introducing
the couplings $\rho_u$ and $\rho_d$ to $K$ and $\lambda_u$ and $\lambda_d$
to $H$, the Yukawa Lagrangian writes
\footnote{$\tau^2$ is the Pauli matrix $\left(\begin{array}{rr} 0 & -i \cr
i & 0 \end{array}\right)$. The doublets $\tilde K
\equiv i\tau^2 K^\ast$ and $\tilde H\equiv i\tau^2 H^\ast$ are isomorphic
to $K$ and $H$.}
\begin{eqnarray}
{\cal L}_{Yukawa}= &+&\rho_d
\left(\begin{array}{cc}\overline{u_L}\
\overline{d_L}\end{array}\right) K
\, d_R
- \rho_u
\left(\begin{array}{cc}\overline{u_L}\
\overline{d_L}\end{array}\right)
(i\tau^2 K^\ast)
\, u_R\cr
&+&\lambda_d
\left(\begin{array}{cc}\overline{u_L}\
\overline{d_L}\end{array}\right) H
\, d_R
+\lambda_u \left(\begin{array}{cc}\overline{u_L}\
\overline{d_L}\end{array}\right)
(i\tau^2 H^\ast)
\, u_R\cr
&+&   h.c.,
\label{eq:genyuk1}
\end{eqnarray}
which gives, explicitly,
\begin{eqnarray}
\hskip -.6cm{\cal L}_{Yukawa} =
&-&\left[\delta_1\frac{v}{\sqrt{2}\mu^3}(\bar u u+\bar d d)
+\kappa_{12}\frac{\sigma}{\sqrt{2}\nu^3}(\bar u u-\bar d d)\right]{\mathfrak s}^0
 -\left[\delta_{12}\frac{v}{\sqrt{2}\mu^3}(\bar u u+\bar d d)
+\delta_2\frac{\sigma}{\sqrt{2}\nu^3}(\bar u u-\bar d d)\right]{\mathfrak s}^3\cr
& \cr
 &+&\left[
 \delta_1\frac{v}{\sqrt{2}\mu^3}
\Big(\bar u\gamma_5 d\, {\mathfrak p}^- +\bar d\gamma_5 u\,
{\mathfrak p}^+ +(\bar u \gamma_5 u -\bar d \gamma_5 d)\,{\mathfrak
p}^3\Big)
+\kappa_{12}\frac{\sigma}{\sqrt{2}\nu^3} \Big(\bar d u\, {\mathfrak p}^+-\bar u d\, {\mathfrak
p}^- +(\bar u \gamma_5 u + \bar d\gamma_5 d)\,{\mathfrak p}^3\Big)
\right]
\cr
& \cr
&-&\left[
\delta_{12}\frac{v}{\sqrt{2}\mu^3}\Big(\bar d\gamma_5 u\, {\mathfrak s}^+ -\bar u\gamma_5
d\,{\mathfrak s}^- -(\bar u\gamma_5 u - \bar d\gamma_5 d)\,{\mathfrak
p}^0\Big)
+ \delta_2\frac{\sigma}{\sqrt{2}\nu^3}\Big(\bar d u\, {\mathfrak
s}^+ + \bar u d\, {\mathfrak s}^--(\bar u\gamma_5 u + \bar d\gamma_5
d)\,{\mathfrak p}^0\Big)
\right].\cr
&&
\label{eq:genyuk2}
\end{eqnarray}
In (\ref{eq:genyuk1}) and (\ref{eq:genyuk2}) the signs have been set such that
for positive $<{\mathfrak s}^0>$ and $<{\mathfrak s}^3>$,
the fermion masses are positive for positive
$\rho_{u,d}$ and $\lambda_{u,d}$ (given that a fermion mass term is of the
form $-m\bar\psi \psi$).
We introduced in (\ref{eq:genyuk2}) the  parameters with dimension $[mass]^2$ 
\begin{eqnarray}
\delta_1 &=& \displaystyle\frac{\rho_u +
\rho_d}{2}\,\frac{\sqrt{2}\mu^3}{v}  ,\cr
&&\cr
\kappa_{12} &=& \displaystyle\frac{\rho_u -
\rho_d}{2}\,\frac{\sqrt{2}\nu^3}{\sigma}  ,\cr
&&\cr
\delta_{12} &=&
\displaystyle\frac{\lambda_u+\lambda_d}{2}\,\frac{\sqrt{2}\mu^3}{v}  ,\cr
&&\cr
\delta_2 &=& \displaystyle\frac{\lambda_u-\lambda_d}{2}\,\frac{\sqrt{2}\nu^3}{\sigma}.
\label{eq:params}
\end{eqnarray}

\subsection{Fermion masses}
\label{subsec:fermass}

We define the two quantum Higgs fields $\varsigma$ and $\xi$ by shifting the scalar
fields  ${\mathfrak s}^0$ and ${\mathfrak s}^3$ 
occurring respectively  in the Higgs doublets
 $K$ and $H$ (see (\ref{eq:K}),(\ref{eq:H})) according to
\begin{equation}
{\mathfrak s}^0 = <{\mathfrak s}^0> + \varsigma, \quad
{\mathfrak s}^3 = <{\mathfrak s}^3> + \xi.
\label{eq:defhxi}
\end{equation}
The two VEV's (given in (\ref{eq:K}) and (\ref{eq:H}))
 contribute to the fermion masses according to
\begin{equation}
m_u=\rho_u<{\mathfrak s}^0> +\lambda_u<{\mathfrak s}^3>
=\frac{v\rho_u + \sigma\lambda_u}{\sqrt{2}},\quad
m_d=\rho_d<{\mathfrak s}^0> +\lambda_d<{\mathfrak s}^3>
=\frac{v\rho_d + \sigma\lambda_d}{\sqrt{2}}.
\label{eq:mumd}
\end{equation}

Additional remarks concerning fermion masses are written in subsection
\ref{subsec:fmasslowen}.

\section{The low energy limit}
\label{section:lowen}

At low energy we use the one-to-one correspondence between $K,H$ and

\vbox{
\begin{eqnarray}
{\mathfrak K}
=\frac{1}{\sqrt{2}}\frac{v}{\mu^3}\left(\begin{array}{c}
\phi^1-i\phi^2 \cr
-(\phi^0+\phi^3)\end{array}\right)
&=&\frac{v\sqrt{2}}{\mu^3}\left(\begin{array}{c}
\bar d \gamma_5 u \cr
-\frac12(\bar u u + \bar d d) -\frac12(\bar u\gamma_5 u - \bar d \gamma_5 d)
\end{array}\right)
\equiv\left(\begin{array}{c} {\mathfrak k}^1-i{\mathfrak
k}^2\cr-({\mathfrak k}^0+{\mathfrak k}^3)\end{array}\right),\cr
&& \cr
<\bar u u + \bar d d> &=& \mu^3,\cr
&& \cr
{\mathfrak H}
=\frac{1}{\sqrt{2}}\frac{\sigma}{\nu^3}\left(\begin{array}{c}
\xi^1-i\xi^2\cr
-(\xi^0+\xi^3)\end{array}\right)
&=&\frac{\sigma\sqrt{2}}{\nu^3}\left(\begin{array}{c}
\bar d  u \cr
-\frac12(\bar u\gamma_5 u + \bar d\gamma_5 d) -\frac12(\bar u u - \bar d  d)
\end{array}\right)
\equiv\left(\begin{array}{c} {\mathfrak h}^1-i{\mathfrak
h}^2\cr-({\mathfrak h}^0+{\mathfrak h}^3)\end{array}\right),\cr
&& \cr
<\bar u u - \bar d d> &=& \nu^3,\cr
&&
\label{eq:compdoub}
\end{eqnarray}
}

that has been established in \cite{Machet1} and identify accordingly
\begin{equation}
(\mathfrak{s}^0, \mathfrak{p}^1, \mathfrak{p}^2, \mathfrak{p}^3) \simeq
(\mathfrak{k}^0, \mathfrak{k}^1, \mathfrak{k}^2, \mathfrak{k}^3),\quad
(\mathfrak{p}^0, \mathfrak{s}^1, \mathfrak{s}^2, \mathfrak{s}^3) \simeq
(\mathfrak{h}^0, \mathfrak{h}^1, \mathfrak{h}^2, \mathfrak{h}^3).
\label{eq:ident}
\end{equation}

\subsection{Rewriting Yukawa couplings}
\label{subsec:newyuk}

The first consequence of this correspondence is that, defining
\begin{equation}
m_{12}^2 = \kappa_{12}+ \delta_{12},
\label{eq:defm12}
\end{equation}
and expressing the bilinear quark operators in (\ref{eq:genyuk2})
 in terms of the components
$({\mathfrak s}^0, \vec {\mathfrak p}),({\mathfrak p}^0,\vec {\mathfrak s})$ of
$K$ and $H$, the Yukawa couplings (\ref{eq:genyuk2})  rewrite
\begin{equation}
{\cal L}_{Yukawa}^{eff}=-\delta_1\, K^\dagger K - \frac12m_{12}^2\,
(K^\dagger H + H^\dagger K) -\delta_2\, H^\dagger H,
\label{eq:lowenyuk1}
\end{equation}
or, indifferently, since renormalizability is not an issue at low energy,
as  a sum of 4-fermion interactions
\begin{equation}
{\cal L}_{Yukawa}^{eff}=-\delta_1\, \mathfrak{K}^\dagger \mathfrak{K}
- \frac12m_{12}^2\,
(\mathfrak{K}^\dagger \mathfrak{H} + \mathfrak{H}^\dagger \mathfrak{K})
-\delta_2\, \mathfrak{H}^\dagger \mathfrak{H}.
\label{eq:lowenyuk2}
\end{equation}

This bosonised form of the Yukawa couplings, only valid at low energy, will
be later added to the scalar potential $V(K,H)$ to define the low energy effective
potential $V_{eff}(K,H)$ (see subsection \ref{subsec:effpot}).

\subsection{PCAC and the Gell-Mann-Oakes-Renner relation}
\label{subsec:pimass}

Kinetic terms together with Yukawa couplings 
include in particular
\begin{equation}
(\partial_\mu K)^\dagger \partial^\mu K -\delta_1 K^\dagger K
- \frac12m_{12}^2\,
({K}^\dagger {H} + {H}^\dagger {K})
+ (\partial_\mu H)^\dagger \partial^\mu H
-\delta_2\, {H}^\dagger {H} +\ldots
\label{eq:kinyuk}
\end{equation}
and we now raise the issue whether, at low energy,
the charged components of $K$ can be
identified with the charged pions. As we shall see in subsection
\ref{subsec:pieta} below, both $\delta_2$ and $m_{12}^2$ have to vanish: the
first to ensure that the breaking of the weak $SU(2)_L$ is accompanied by
three true Goldstone bosons, and the second to ensure that the ${\mathfrak
p}^0-{\mathfrak p}^3$ system does not exhibit a tachyonic state.
Eq.~(\ref{eq:kinyuk}) reduces then to  standard kinetic terms  for 
unmixed doublets. Furthermore, the scalar potential will be chosen in such
a way that the three pseudoscalar bosons inside $K$ are Goldstone bosons in
the absence of Yukawa couplings. So, due to the mass term proportional to
$\delta_1$, the three ``pions'' inside $K$ get a mass $m$ at
the simple condition that $\delta_1= \frac12 m^2$.  

Owing to the Partially Conserved Axial Current (PCAC) hypothesis
\cite{Dashen}\cite{Lee}
\begin{equation}
i(m_u+m_d)\,\bar u \gamma_5 d= \sqrt{2} f_\pi m_\pi^2\, \pi^+,
\label{eq:PCAC}
\end{equation}
which identifies the interpolating pion field with a bilinear quark
operator, and to the corresponding Gell-Mann-Oakes-Renner relation \cite{GMOR}
\begin{equation}
(m_u+m_d)<\bar u u + \bar d d> =2f_\pi^2 m_\pi^2,
\label{eq:GMOR}
\end{equation}
${\mathfrak p}^+ \equiv {\mathfrak p}^1+i{\mathfrak p}^2 =
\frac{v\sqrt{2}}{\mu^3}\bar d\gamma_5 u$ as it is defined in
(\ref{eq:lowenyuk1}) and (\ref{eq:ident}) can be identified at low energy
with
\begin{equation}
{\mathfrak p}^\pm \simeq -\frac{iv}{f_\pi} \pi^\pm.
\label{eq:ppi1}
\end{equation}
So, the kinetic terms $ (\partial_\mu K)^\dagger \partial^\mu K$,
which contain in particular $ \partial_\mu {\mathfrak p}^+ \partial
^\mu {\mathfrak p}^- \equiv  \left( \partial_\mu {\mathfrak p}^1 \partial
^\mu {\mathfrak p}^1  +  \partial_\mu {\mathfrak p}^2 \partial
^\mu {\mathfrak p}^2\right)$, will be normalized in the standard way if 
\begin{equation}
v=f_\pi,
\label{eq:v}
\end{equation}
such that
\begin{equation}
{\mathfrak p}^\pm \simeq -i\pi^\pm.
\label{eq:ppi}
\end{equation}
Then, the term proportional to $\delta_1$ in (\ref{eq:kinyuk}) is a
suitable pion mass terms if
\begin{equation}
\delta_1 =  m_\pi^2.
\label{eq:delta1}
\end{equation}
Going back to the definition of $\delta_1$ in (\ref{eq:params}) and using
(\ref{eq:v}) and (\ref{eq:GMOR}), (\ref{eq:delta1}) corresponds to
\begin{equation}
\rho_u + \rho_d = \frac{m_u+m_d}{\sqrt{2}f_\pi}.
\label{eq:rou+rod}
\end{equation}
Since $f_\pi \ll m_W$, (\ref{eq:v}) plugged into (\ref{eq:mw}) entails
\begin{equation}
\sigma \approx \frac{2 m_W}{g},
\label{eq:sigma}
\end{equation}
which shows that the $\vec W$'s get their mass essentially from the VEV of
${\mathfrak s}^3$. 
The ratio of the VEV's of the two Higgs doublets comes out accordingly as
\begin{equation}
\tan\beta = \frac{<{\mathfrak s}^3>}{<{\mathfrak s}^0>}=
 \frac{\sigma/\sqrt{2}}{v/\sqrt{2}} \approx \frac{2m_W}{gf_\pi}
\approx 2856.
\label{eq:rapvev}
\end{equation}
They correspond respectively to the weak ($m_W$) and chiral ($f_\pi$)
scale. Both scales can now coexist, unlike in the genuine
Glashow-Salam-Weinberg model where the parity-transformed $H$ of the Higgs
doublet $K$ is missing.

Eqs.~(\ref{eq:mumd}) and (\ref{eq:sigma})
then determine  $\lambda_u$ and $\lambda_d$
\begin{equation}
(\lambda_u+\lambda_d) = \frac{g}{2\sqrt{2}m_W}(m_u+m_d),\quad
(\lambda_u-\lambda_d) =
\frac{g}{\sqrt{2}m_W}\left((m_u-m_d)-\frac{f_\pi}{\sqrt{2}}(\rho_u-\rho_d)\right),
\label{eq:lambda1}
\end{equation}
that is
\begin{equation}
\lambda_u=g\,\frac{3 m_u-m_d-2\sqrt{2}f_\pi(\rho_u-\rho_d)}{4\sqrt{2}\,m_W},\quad
\lambda_d=g\,\frac{3 m_d-m_u+2\sqrt{2}f_\pi(\rho_u-\rho_d)}{4\sqrt{2}\,m_W},
\label{eq:lambda2}
\end{equation}
in terms of $\rho_u - \rho_d$ which is, at this point, still undetermined.

\subsection{Goldstones and pseudo-Goldstones}
\label{subsec:pieta}

\subsubsection{The charged Goldstones of the broken $\boldsymbol{SU(2)_L}$}
\label{subsub:xigolds}

Since $<{\mathfrak s}^3>$ provides most of the mass of the $\vec W$'s,
the charged Goldstone bosons of the broken $SU(2)_L$ weak symmetry are, to
a very good approximation, the excitations of ${\mathfrak s}^3$ by the
generators $T^+_L$ and $T^-_L$, that is $\mathfrak{s}^+$ and
$\mathfrak{s}^-$ $\in H$.

However, the $SU(2)_L$ invariant Yukawa couplings that need to be
introduced to provide fermions with ``soft'' masses also give, at low
energy, among other couplings,
a ``soft'' mass to  $\mathfrak{s}^+$ and $\mathfrak{s}^-$
through the term proportional to $\delta_2$.
The situation for  $\mathfrak{s}^+$ and $\mathfrak{s}^-$ is different
from that of the pions which can 
become pseudo-Goldstone bosons and stay as physical particles.
The spontaneously broken $SU(2)_L$ symmetry requires  true Goldstones,
which can only go along with
\begin{equation}
\delta_2=0,
\label{eq:delta2}
\end{equation}
which is accordingly a side-effect of weak symmetry breaking.
Looking at (\ref{eq:params}), one could think that $\nu^3 \equiv <\bar u
u-\bar d d>=0$ could be a solution to $\delta_2=0$. However, we shall see
later in subsection \ref{subsec:condensates} that $<\bar u u>$ must be different from
$<\bar d d>$ as a trigger of both weak and custodial symmetry breaking.
So, (\ref{eq:delta2}) entails
\begin{equation}
 \lambda_u = \lambda_d =
\frac{g}{4\sqrt{2}\,m_W}(m_u+m_d).
\label{eq:lambda}
\end{equation}

By (\ref{eq:lambda1}), (\ref{eq:lambda}) determines
\begin{equation}
\rho_u-\rho_d=\frac{\sqrt{2}(m_u-m_d)}{f_\pi},
\label{eq:rou-rod}
\end{equation}
and, combined with (\ref{eq:rou+rod}),
\begin{equation}
\rho_u=\frac{3m_u-m_d}{2\sqrt{2}f_\pi},\quad \rho_d=\frac{3m_d-m_u}{2\sqrt{2}f_\pi}.
\label{eq:rourod}
\end{equation}

\subsubsection{The $\boldsymbol{{\mathfrak p}^3-{\mathfrak p}^0}$ system}
\label{subsub:pieta}

The $(\mathfrak{p}^3,\mathfrak{p}^0)$ or $(\mathfrak{k}^3,\mathfrak{h}^0)$ or, equivalently $(\pi^0,\eta)$ system gets
endowed by the Yukawa couplings with a mass matrix
\begin{equation}
\frac12\left(\begin{array}{cc}
2\delta_1 & m_{12}^2\cr
 m_{12}^2 & 2\delta_2
\end{array}\right).
\end{equation}
However, since $\delta_2$ has been fixed to zero in  subsection
\ref{subsub:xigolds}, this system now exhibits a
tachyonic state unless
\begin{equation}
m_{12}^2=0 \Leftrightarrow
(\rho_u-\rho_d)\frac{\nu^3}{\sigma}=-(\lambda_u+\lambda_d)\frac{\mu^3}{v}
\Leftrightarrow \frac{m_u-m_d}{m_u+m_d}=-\frac12\frac{\mu^3}{\nu^3}
\equiv -\frac12 \frac{<\bar u u + \bar d d>}{<\bar u u-\bar d d>}, 
\label{eq:m12vanish}
\end{equation}
in which we have used (\ref{eq:defm12}).(\ref{eq:params}),
(\ref{eq:lambda}), (\ref{eq:rourod}) and the
definitions of $\mu^3$ and $\nu^3$ that were introduced in
(\ref{eq:compdoub}).

Eq.~(\ref{eq:m12vanish}) is equivalent to
\footnote{$<\bar d d>$ vanishes for $m_d=3 m_u$. We shall see in subsection
\ref{subsec:htoq} that this is also the condition for the $u$ quark to couple to
the ``standard'' Higgs boson $\xi$ like in the Glashow-Salam-Weinberg
model.}
\begin{equation}
\frac{<\bar d d>}{<\bar u u>}= \frac{3m_u-m_d}{m_u-3m_d}.
\label{eq:rapdduu}
\end{equation}
When this is realized,
${\mathfrak p}^0$  is a true Goldstone and ${\mathfrak p}^3$ keeps
its mass $m_\pi^2$. They do not mix.
This fits the picture of  ${\mathfrak p}^0$  being the third Goldstone
boson of the broken $SU(2)_L$ symmetry, and  ${\mathfrak p}^3$ being the
neutral member of the triplet of pseudo-Goldstone bosons of the broken
chiral symmetry $SU(2)_L \times SU(2)_R$ down to the diagonal $SU(2)$.
Other considerations concerning  symmetries will be given in section
\ref{section:symmetries}. 

\subsubsection{No scalar-pseudoscalar coupling}
\label{subsub:noscps}

Yukawa couplings are seen on (\ref{eq:genyuk2}) to potentially generate
couplings between charged scalars, for example ${\mathfrak
s}^-=\frac{\sigma}{\sqrt{2}\nu^3}\bar d u$ and pseudoscalars, for example
${\mathfrak p}^+$. It is the second important effect of the condition
$m_{12}^2 \equiv \delta_{12}+\kappa_{12}=0$ obtained in subsection
\ref{subsub:pieta} to cancel these transitions.

\subsubsection{The unitary gauge. Leptonic decays of pions}
\label{subsub:pilep}

In the unitary gauge the crossed couplings between the $\vec W$ gauge
bosons and the (derivative of the) $SU(2)_L$  Goldstone bosons ${\mathfrak
p}^0, {\mathfrak s}^+, {\mathfrak s}^-$ are canceled, which leaves untouched
the similar couplings between $\vec W$ and the three pions. Their
proportionality to $v = f_\pi$  yields in
particular leptonic decays of pions in agreement with the standard PCAC
calculation. 

\subsection{Fermion masses versus the low energy effective Lagrangian}
\label{subsec:fmasslowen}

Fermions receive their masses from the VEV's of the two Higgs doublets $K$
and $H$. From (\ref{eq:mumd}) and the values of the parameters that have
been determined (see also section \ref{section:paramsummary} below), it
appears that $<{\mathfrak s}^3> \in H$ contribute to the $u$ and
$d$ masses by the same amount
$\frac{\sigma\lambda_u}{\sqrt{2}}=\frac{\sigma\lambda_d}{\sqrt{2}} =
\frac{m_u+m_d}{4}$. Then, $<{\mathfrak s}^0> \in K$ contributes to the $u$
mass by $\frac{v\rho_u}{\sqrt{2}}= \frac{3m_u-m_d}{4}$ and to the $d$ mass
by $\frac{v\rho_d}{\sqrt{2}}= \frac{3m_d-m_u}{4}$.

The second point is the inadequacy to calculate quark
masses from  the low energy effective expression
(\ref{eq:lowenyuk1}) of the Yukawa couplings and its set of parameters
determined by low energy considerations.
When plugged into (\ref{eq:lowenyuk1}) the conditions $\delta_2=0$ and
 $m_{12}^2\equiv \delta_{12}+\kappa_{12}=0$ demonstrated respectively in
(\ref{eq:delta2}) and in (\ref{eq:m12vanish}) entail that quark masses 
come  from  the sole Higgs doublet $K$, by  $-\delta_1 K^\dagger K$.
Going back to quark
fields and writing it for example as the product
 $-\delta_1 K^\dagger {\mathfrak K}$ of scalar fields $K$ times their
equivalents in terms of bilinear quark operators $\mathfrak K$,
which respects renormalizability, ${\cal L}_{Yukawa}^{eff}$ does, 
through quark-antiquark condensation,  generate quark masses. They however come out
as $-\delta_1\frac{v^2}{2\mu^3}(\bar u u+\bar d d)
=-\frac{m_u+m_d}{4} (\bar u u+\bar d d)$, which  is different
from the masses obtained  from the original Lagrangian (\ref{eq:genyuk2})
\begin{equation}
-\delta_1 \frac{v}{\sqrt{2}\mu^3}(\bar u u+\bar d
d)<{\mathfrak s}^0> +\delta_{12}\left(\frac{\sigma}{\sqrt{2}\nu^3}(\bar u
u-\bar d d) <{\mathfrak s}^0>-\frac{v}{\sqrt{2}\mu^3}(\bar u u+\bar d
d)<{\mathfrak s}^3> \right);
\label{eq:fmass}
\end{equation}
using the expression for $\delta_{12}$ deduced from
(\ref{eq:params}) and (\ref{eq:lambda1}), the genuine Lagrangian
 (\ref{eq:fmass}) yields the mass terms
\begin{equation}
\begin{split}
-\delta_1 \frac{v^2}{2\mu^3}(\bar u u+\bar d d)
+\delta_{12}\frac{v\sigma}{2}\left(\frac{\bar u u-\bar d d}{\nu^3}
-\frac{\bar u u+\bar d d}{\mu^3} \right)
& \stackrel{(\ref{eq:sumparams})}{=}
 -\frac{f_\pi^2m_\pi^2}{2}\frac{\bar u u+\bar d d}{\mu^3}
+\frac{f_\pi^2m_\pi^2}{2}\left(\frac{\bar u u-\bar d d}{\nu^3}-\frac{\bar
u u+\bar d d}{\mu^3}\right)\cr
&\hskip -4.5cm \stackrel{(\ref{eq:sumparams})}{=}
 -\underbrace{\frac14(m_u+m_d)(\bar u u+\bar d d)}_{\text{from}\ \delta_1}
 -\underbrace{\frac12(m_u-m_d)(\bar u u-\bar d d)}_{\text{from}\ \kappa_{12}} 
-\underbrace{\frac14(m_u+m_d)(\bar u u+\bar d
d)}_{\text{from}\ \delta_{12}}.
\end{split}
\label{eq:massterms}
\end{equation}
 In (\ref{eq:massterms}), unlike in ${\cal L}_{Yukawa}^{eff}$, the terms proportional to $\delta_{12}$
do not vanish because the bilinear fermion operators do not reduce to their
low energy VEV's $<\bar u u-\bar d d>=\nu^3, <\bar u u+\bar d d>=\mu^3$.
Furthermore, even if $m_u$ is set equal to $m_d$, the part proportional to
$\delta_{12}$, which describes  $H-K$ interplay,
 contributes to quark masses  as much as the one proportional
 to $\delta_1$ which comes from $K$ alone.
Therefore, neither the effective Lagrangian ${\cal L}^{eff}_{Yukawa}$ nor
 the ``low energy truncation'' of the model, that includes
only one Higgs doublet, $K$, can correctly account for fermion masses
(nor, of course, for the masses of the gauge bosons, problem which led to
``technicolor'' models \cite{Susskind}).
 ${\cal L}_{Yukawa}^{eff}$ we shall accordingly only use
to deal with low energy physics of scalars and pseudoscalars, in particular
to build the effective scalar potential $V_{eff}$ in subsection
\ref{subsec:effpot}.

\section{Summary of the parameters}
\label{section:paramsummary}
By low energy considerations, we have determined the following parameters,
introduced in particular in (\ref{eq:genyuk1}) and (\ref{eq:params}):

\vbox{
\begin{eqnarray}
&&\rho_u=\frac{3m_u-m_d}{2\sqrt{2}f_\pi},\quad
\rho_d=\frac{3m_d-m_u}{2\sqrt{2}f_\pi},\quad
\lambda_u=\lambda_d= \frac{g(m_u+m_d)}{4\sqrt{2}m_W},\cr
&& \cr
&& \delta_1= m_\pi^2,\quad
\delta_{12}=-\kappa_{12}=\frac{gf_\pi m_\pi^2}{2m_W},\quad \delta_2=0,\cr
&& \cr
&& (m_u+m_d)<\bar u u + \bar d d> \stackrel{(\ref{eq:GMOR})}{=}
 2f_\pi^2m_\pi^2,\quad
(m_u-m_d)<\bar u u-\bar d d> \stackrel{(\ref{eq:m12vanish})}{=}
- f_\pi^2 m_\pi^2,\cr
&& \cr
&& v\equiv \sqrt{2}<{\mathfrak s}^0>=f_\pi, \quad \sigma\equiv
\sqrt{2}<{\mathfrak s}^3>=\frac{2m_W}{g}.
\label{eq:sumparams}
\end{eqnarray}
}

These should be plugged into the renormalizable form (\ref{eq:genyuk2}) of the
Yukawa Lagrangian. Note that, unlike its low energy avatar
(\ref{eq:lowenyuk1}), it depends
on $\kappa_{12}$ and $\delta_{12}=-\kappa_{12}$, and not on
$m_{12}^2=0$.

\section{The scalar potential}
\label{section:pot}

\subsection{A $\boldsymbol{U(2)_L \times U(2)_R}$ invariant potential}
\label{subsec:pot}

We shall consider a quartic $U(2)_L \times U(2)_R$  invariant potential
\begin{equation}
V(K,H) = -\frac{m_H^2}{2}\, (K^\dagger K + H^\dagger H)
         +\frac{\lambda_H}{4}\,\Big((K^\dagger K)^2 + (H^\dagger H)^2\Big),
\label{eq:pot}
\end{equation}
which thus decomposes into two independent potentials, one for $K$ and one
for $H$.

This is possible because (see \cite{Machet1})
$K$ and $H$ are stable by both  $SU(2)_L$ and
$SU(2)_R$ and transform into each other by $U(1)_L$ and $U(1)_R$ (with the
appropriate signs).
This last symmetry dictates in particular the equality
of the couplings (quadratic and quartic) for the two doublets.

$SU(2)_L$ breaking by $v\not=0$ and $\sigma \not=0$ generates three
Goldstone bosons in each Higgs multiplet: $\vec{\mathfrak p}\in K$,
the pseudoscalar singlet ${\mathfrak p}^0$ and the two charged scalars
${\mathfrak s}^\pm \in H$.
This also fits the scheme according to which  $v\not=0$ and $\sigma\not=0$
spontaneously break the chiral $U(2)_L \times U(2)_R$ down to $U(1) \times U(1)_{em}$
(see \cite{Machet1});
there, too,  six Goldstones are generated. The pseudoscalar triplet
$\vec{\mathfrak p}\in K$ gets a small mass from the $SU(2)_L$
invariant Yukawa couplings while the pseudoscalar singlet ${\mathfrak p}^0$ and
the two charged scalars ${\mathfrak s}^\pm \in H$ must be protected from
this since they are also the three Goldstones to be eaten by the $\vec W$
gauge bosons (see section \ref{section:lowen}).
 ${\mathfrak p}^0$ plays a double role in that is also the Goldstone of the
breaking of $U(1)_L \times U(1)_R$ down to the diagonal $U(1)$, which, at
the level of the algebra, is related to parity breaking. 

$v\not=0$ is associated with $<\bar u u + \bar d d> \not=0$,
responsible for the breaking of $SU(2)_L \times SU(2)_R$ down to $SU(2)$
with the three pions as (pseudo)-Goldstone bosons, while $\sigma \not=0$ is
associated with $<\bar u u-\bar d d>\not=0$ which
is also responsible for the breaking of the custodial $SU(2)$ into $U(1)_{em}$
and of the $\vec W$ mass.

Our choice for the potential amounts to requesting that, in the absence of
Yukawa couplings, all fields are Goldstones but for the two Higgs bosons.

In the most general potential for two Higgs doublets the following terms
have accordingly been discarded:

$\bullet$\quad $(m^2 K^\dagger H +h.c)$, with $m \in {\mathbb C}$
would mediate in particular transitions between scalars and pseudoscalars
that should not occur classically;

$\bullet$\quad $\lambda_4(K^\dagger K)(K^\dagger H) + h.c.$,
$\lambda_5(H^\dagger H)(K^\dagger H_2) + h.c.$
with $\lambda_4, \lambda_5 \in {\mathbb C}$ would also mediate 
unwanted classical transitions between scalars and pseudoscalars;

$\bullet$\quad $\lambda_3(K^\dagger H)^2 + h.c.$ with $\lambda_3 \in {\mathbb C}$
would in particular  contribute to the mass of the neutral pion and not
to that of the charged pions. Such a classical $\pi^+-\pi^0$ mass
difference which is not electromagnetic nor due to $m_u \not= m_d$ is
unwelcome;

$\bullet$\quad $\lambda_1(K^\dagger K)(H^\dagger H)$, 
$\lambda_2(K^\dagger H)(H^\dagger K)$,
with $\lambda_1, \lambda_2 \in {\mathbb R}$ would also spoil the
Goldstone nature of the pions and $\eta$, the first because of terms
proportional to $<{\mathfrak s}^3>^2 \vec\pi^2$ and $<{\mathfrak s}^0>^2
\eta^2$, the second because of terms proportional to $<{\mathfrak s}^0>^2
\eta^2, <{\mathfrak s}^3>^2 {\pi^0}^2$ and $<{\mathfrak s}^0><{\mathfrak
s}^3> \pi^0 \eta$.

\subsection{The low energy effective potential}
\label{subsec:effpot}

At low energy, the renormalizable $V(K,H)$ is supplemented by $(-1)\times$
the bosonised form of the Yukawa Lagrangian (\ref{eq:lowenyuk1}).
This yields the effective potential
\begin{eqnarray}
V_{eff}(K,H) &=& V(K,H)
+\delta_1\, K^\dagger K + \frac12 m_{12}^2\, (K^\dagger H
+ H^\dagger K) +\delta_2\, H^\dagger H\cr
&& \hskip -2cm = -\frac{m_H^2}{2}\, (K^\dagger K + H^\dagger H)
         +\frac{\lambda_H}{4}\,\Big((K^\dagger K)^2 + (H^\dagger H)^2\Big)
+\delta_1\, K^\dagger K + \frac12 m_{12}^2\, (K^\dagger H
+ H^\dagger K) +\delta_2\, H^\dagger H.\cr
&&
\label{eq:effpot1}
\end{eqnarray}

It is further simplified since we have shown that $\delta_2=0$ and $m_{12}^2=0$ 
(see (\ref{eq:delta2}) and (\ref{eq:lambda}) in section \ref{section:lowen})
  and $V_{eff}$ accordingly reduces to
\begin{equation}
V_{eff}(K,H) =
-\frac{m_H^2-2 m_\pi^2}{2}\, K^\dagger K -\frac{m_H^2}{2}\, H^\dagger H
         +\frac{\lambda_H}{4}\,\Big((K^\dagger K)^2 + (H^\dagger H)^2\Big).
\label{eq:effpot2}
\end{equation}

Last, to suitably reproduce the $\vec\pi$  and $\vec W$ masses, we know that it should
have a minimum at values of $v$ and $\sigma$ given by (\ref{eq:v}) and
(\ref{eq:sigma}). The two equations $\frac{\partial V_{eff}}{\partial
{\mathfrak s}^0}\Big|_{<{\mathfrak s}^0>=\frac{f_\pi}{\sqrt{2}}}=0$ and
 $\frac{\partial V_{eff}}{\partial
{\mathfrak s}^3}\Big|_{<{\mathfrak s}^3>=\frac{\sqrt{2}m_W}{g}}=0$ yield
respectively
$m_H^2= \lambda_H <{\mathfrak s}^0>^2 + 2 m_\pi^2$ and $m_H^2 = \lambda_H
<{\mathfrak s}^3>^2$ such that
\begin{equation}
\lambda_H=\frac{2 m_\pi^2}{<{\mathfrak s}^3>^2-<{\mathfrak s}^0>^2}
\approx \frac{2 m_\pi^2}{<{\mathfrak s}^3>^2}\left(1+\frac{<{\mathfrak
s}^0>^2}{<{\mathfrak s}^3>^2}\right)
= \frac{g^2 m_\pi^2}{m_W^2}\left(1+\frac{g^2 f_\pi^2}{4 m_W^2}\right),
\label{eq:lambdaeff}
\end{equation}
which puts it definitely in the perturbative regime.
It is because of the presence of $m_\pi^2$ that $\lambda_H$ is different
from zero. $m_\pi\not=0$ keeps accordingly the theory away from
instability.

\subsection{The masses of the two Higgs bosons $\boldsymbol{\varsigma}$
 and $\boldsymbol{\xi}$} \label{subsec:hmass}

Since the effective scalar potential is now fully determined, one can
calculate the masses  of the two Higgs bosons $\varsigma$ and $\xi$ defined in
(\ref{eq:defhxi}), which do not mix.  One gets
\begin{eqnarray}
m_\xi &=& <{\mathfrak s}^3> \sqrt{\lambda_H}  \approx \sqrt{2}\, m_\pi,\cr
m_\varsigma  &=&  <{\mathfrak s}^0> \sqrt{\lambda_H}
=m_\xi \frac{<{\mathfrak s}^0>}{<{\mathfrak s}^3>} \approx  m_\pi\frac{
gf_\pi}{\sqrt{2}\,m_W} \approx 68\,KeV,
\label{eq:mhiggs}
\end{eqnarray}
In particular, their ratio is that of the two VEV's
\begin{equation}
\frac{m_\xi}{m_\varsigma} = \frac{<{\mathfrak s}^3>}{<{\mathfrak
s}^0>}=\frac{2 m_W/g}{f_\pi}
\end{equation}
which is also the ratio of the two scales involved in this 1-generation standard
model, the weak scale $\simeq m_W$ and the chiral scale $\simeq f_\pi$.
The masses are small and justify {\em a posteriori} our low energy
treatment of the scalar effective potential.

The composition of the two Higgs doublets is accordingly as follows.
Inside the ``chiral'' doublet $K$ one finds 3 pions and the very light
scalar Higgs boson $\varsigma$. As was shown in \cite{Machet1}, they correspond
respectively to a triplet and a singlet of the custodial $SU(2)$ symmetry.
Inside the ``weak'' doublet $H$, one finds the three Goldstones of the broken
$SU(2)_L$ weak symmetry, the neutral pseudoscalar $SU(2)$ singlet
and two charged scalars inside the $SU(2)$ triplet. The third component of this
triplet is the second scalar Higgs boson $\xi$ with mass $\approx m_\pi$.
Note that the four particles $(\vec\pi, \xi)$ with mass $m_\pi$ do not lie
together inside the same $SU(2)_L$ doublet, nor do the three $SU(2)_L$ Goldstones
and the very light Higgs boson $\varsigma$.

\subsubsection{The roles of $\boldsymbol{m_W}$ and $\boldsymbol{m_\pi}$}
\label{subsub:wpi}

In our rebuilding of Standard Model with only one generation, we find that
the masses of the two Higgs bosons are both proportional to $m_\pi$ and
small. But they are not small in the same way. If $m_\pi$ is replaced by
the mass of some heavier bound state $m \leq \sqrt{2}m_W/g \equiv <{\mathfrak
s}^3>\ \approx 168\,GeV$, $m_\varsigma$ will stay very small
$m_\varsigma \leq f_\pi \approx 
93\,MeV$ while $m_\xi$ will grow like the mass of the bound state. 
So, in the case of more generations, the presence of very light Higgs
boson(s) with a mass lower than $100\,MeV$ looks  a robust feature as
a damping effect of the weak scale $m_W$ but  larger masses can
be expected for some others. It would not be a surprise that,
for 3 generations and up to some coefficient, the mass of one of the Higgs
bosons be set by that  of a bound state involving the top quark.

In the present case, the masses of the two Higgs bosons 
vanish at the limit $m_\pi \to 0$, that is, by the GMOR relation
(\ref{eq:GMOR}), either when $<\bar u u+ \bar d d>\to 0$ or when
$(m_u+m_d)\to 0$. Since we have also determined (see (\ref{eq:sumparams}))
that $(m_u-m_d)<\bar u u-\bar d d>$ vanishes with $m_\pi$, this limit
corresponds either to $<\bar u u>=0 =<\bar d d>$ or to $m_u=0=m_d$.

\section{Couplings of the Higgs bosons}
\label{section:couplings}

\subsection{Couplings of Higgs bosons to quarks}
\label{subsec:htoq}

Like for the calculation of fermion masses (see
subsection\ref{subsec:fmasslowen}), the bosonised forms
(\ref{eq:lowenyuk1}) or (\ref{eq:lowenyuk2}) of the Yukawa couplings, 
which are only valid at low energy, is inappropriate to evaluate the couplings of
fermions, in particular those to the Higgs bosons. 
Indeed, plugging into (\ref{eq:lowenyuk1}) or (\ref{eq:lowenyuk2}) 
the relations
 $m_{12}^2 \stackrel{(\ref{eq:defm12})} {\equiv} \delta_{12}+\kappa_{12}=0$
and $\delta_2=0$ 
that we have obtained for the crossed couplings (see (\ref{eq:sumparams}))
from low energy considerations would erroneously leave as the only couplings
of quarks to Higgs bosons the ones
present in $-\delta_1 K^\dagger K$, in which, in particular,
 no coupling exists between the
``quasi-standard'' Higgs boson $\xi$, which belongs to $H$,
 and  quarks. In order to properly
determine these parameters, the original form 
(\ref{eq:genyuk2}) of the Yukawa couplings must instead be used.

Plugging therefore the definition (\ref{eq:defhxi}) into (\ref{eq:genyuk2}) yields
the following couplings of the Higgs bosons $\varsigma$ and $\xi$ to quarks
\begin{equation}
\begin{array}{lll}
& -\varsigma\,(\rho_u \bar u u + \rho_d \bar d d)-\xi\,(\lambda_u \bar u u +
\lambda_d \bar d d) & \cr
=& -\varsigma\,\left(\delta_1\frac{v}{\sqrt{2}\mu^3}(\bar u u+\bar d d)
+\kappa_{12}\frac{\sigma}{\sqrt{2}\nu^3}(\bar u u-\bar d d)\right)
 -\xi\,\left(\delta_{12}\frac{v}{\sqrt{2}\mu^3}(\bar u u+\bar d d)
+\delta_2\frac{\sigma}{\sqrt{2}\nu^3}(\bar u u-\bar d d)\right)&
\end{array}
\label{eq:htoq1}
\end{equation}
which exhibits, of course, the same structure as in (\ref{eq:fmass}) and
which, using the values (\ref{eq:sumparams}) of the parameters,
$\delta_{12}=-\kappa_{12}$ and $\delta_2=0$, yields
\begin{equation}
{\cal L}_{Higgs-quarks}=
-\varsigma\left(\frac{3m_u-m_d}{2\sqrt{2}f_\pi}\,\bar u u + \frac{3m_d-m_u}{2\sqrt{2}f_\pi}\,\bar d
d\right)
-\xi\, \frac{g(m_u+m_d)}{4\sqrt{2} m_W}\, (\bar u u + \bar d d).
\label{eq:htoq2}
\end{equation}
The $\varsigma$ Higgs boson is more strongly coupled to
quarks than $\xi$. Its coupling is still   ``perturbatively''
since $m_u, m_d \ll f_\pi$. It however suggests that, for heavier quarks,
some Higgs boson(s) could strongly couple to hadronic matter.
As far as $\xi$ is concerned,
it looks at first sight  ``quasi-standard'' because it is proportional
to $g m_{quark}/m_W$. It is however not quite
so because  in  the standard case we would have obtained
 $-\frac{g}{\sqrt{2}m_W}(m_u \bar u u+ m_d \bar d d)\,\xi$.
The difference is that, though $u$ and $d$ have different masses,
they now get coupled to $\xi$ with equal strength: unlike in the genuine
Glashow-Salam-Weinberg model, the heavier quark is no more strongly
coupled than the lighter. Taking $m_d=\gamma m_u, \gamma > 1$, the coupling
$-\frac{g(1+\gamma)}{4\sqrt{2}m_W}m_u$ of $\xi$ to $u$ quarks can be very close
to the standard one (it becomes identical for $\gamma=3$, value at which
$<\bar d d>$ vanishes, see footnote in subsection \ref{subsub:pieta}),
 while the one $-\frac{g(1/\gamma + 1)}{4\sqrt{2}m_W}m_d$ of $\xi$ 
to the heavier $d$ is smaller than standard by the factor
$\frac{(1+\gamma)}{4\gamma}$.

\subsection{Couplings of Higgs bosons to gauge bosons}
\label{subsec:htow}

They  arise from the kinetic terms (\ref{eq:kinetic}).
Using (\ref{eq:v}) and (\ref{eq:sigma})), one gets 
\begin{equation}
{\cal L}_{HiggsWW}=\frac{g m_W}{2} W_\mu^2\, \xi
+ \frac{g^2 f_\pi}{4\sqrt{2}} W_\mu^2\, \varsigma.
\label{eq:htow}
\end{equation}
$\xi$ couples accordingly in 
a ``standard'' way $\simeq gm_W$ to  two $W$'s
while the coupling of $\varsigma$, ${\cal O}(g^2 f_\pi)$ is much smaller by a factor
 ${\cal O} (10^{-3})$.

\subsection{Couplings of Higgs bosons to leptons}
\label{subsec:htol}

Yukawa couplings to leptons need introducing four parameters, $\rho_e$ and
$\rho_\nu$ for ${\mathfrak s}^0$ and the quantum Higgs $\varsigma$,
 $\lambda_e$ and $\lambda_\nu$ for ${\mathfrak s}^3$ and the quantum Higgs $\xi$

\vbox{
\begin{eqnarray}
\hskip -1cm{\cal L}_{Yuk-lept}&=&\Big((\rho_\nu \bar\nu \nu +\rho_e \bar e e)\,{\mathfrak s}^0
-(\lambda_\nu \bar\nu \nu + \lambda_e \bar e e)\,{\mathfrak s}^3 \Big)\cr
&+& \left(\frac{\rho_\nu + \rho_e}{2}\Big(\bar \nu\gamma_5 e\, {\mathfrak p}^-
+ \bar e\gamma_5 \nu\, {\mathfrak p}^+ 
+(\bar \nu \gamma_5 \nu -\bar e\gamma_5 e)\,{\mathfrak p}^3\Big) \right)
+\frac{\rho_\nu-\rho_e}{2}\left(\Big(\bar e \nu\, {\mathfrak p}^+
-\bar\nu e\,{\mathfrak p}^-
+(\bar \nu \gamma_5 \nu + \bar e\gamma_5 e)\,{\mathfrak p}^3 \Big)
\right)\cr
&-&\left(\frac{\lambda_\nu+\lambda_e}{2}\Big(\bar e\gamma_5\nu\, {\mathfrak
s}^+ -\bar\nu\gamma_5 e\, {\mathfrak s}^-
-(\bar\nu\gamma_5\nu - \bar e\gamma_5 e)\,{\mathfrak p}^0\Big)\right)
+\frac{\lambda_\nu-\lambda_e}{2}\left(\Big(\bar e\nu\,{\mathfrak s}^+ + \bar\nu
e\,{\mathfrak s}^-
-(\bar\nu\gamma_5\nu + \bar e\gamma_5 e)\,{\mathfrak p}^0 \Big) \right).\cr
&&
\label{eq:lepyuk}
\end{eqnarray}
}

Using again (\ref{eq:v}) and (\ref{eq:sigma} provides the lepton masses
\begin{equation}
m_e= \rho_e\frac{f_\pi}{\sqrt{2}} +\lambda_e\frac{\sqrt{2}m_W}{g},\quad
m_\nu=\rho_\nu\frac{f_\pi}{\sqrt{2}}+\lambda_\nu \frac{\sqrt{2}m_W}{g}.
\label{eq:memnu}
\end{equation}

\subsubsection{The low energy limit}

Let us use again the one-to-one correspondence between the components of
the Higgs multiplets and bilinear quark operators (\ref{eq:compdoub}).
Using PCAC
(\ref{eq:PCAC}) and the Gell-Mann-Oakes-Renner relation (\ref{eq:GMOR}),
 we could relate 
the charged pion fields $\pi^\pm$ and the charged pseudoscalar components
${\mathfrak p}^\pm$ of the Higgs doublet $K$  by (\ref{eq:ppi}).
Yukawa couplings (\ref{eq:lepyuk}) are then seen to trigger, among others,
leptonic decays of charged pions. These come in addition to the ``standard
ones'' obtained from the $W_\mu \partial^\mu \pi$ crossed couplings that
originate from the kinetic terms (\ref{eq:kinetic}) at low energy (see subsection
\ref{subsub:pilep}) and which agree with PCAC usual calculations.

This means that, in a first approximation (and it is not the goal of this
work to go beyond), we should take
\begin{equation}
\rho_\nu \approx  0 \approx \rho_e.
\label{eq:rolep}
\end{equation}
In case observed leptonic pion decay turn out to  differ from PCAC estimates, the
issue could be raised whether (\ref{eq:rolep}) should be revisited.

In relation with (\ref{eq:memnu}) the choice (\ref{eq:rolep})
 leads to a standard coupling of the Higgs boson $\xi$ to leptons,
 proportional to $gm_{lepton}/m_W$,
while the ones of $\varsigma$ vanish (or are extremely close to).

\section{Symmetries again}
\label{section:symmetries}

\subsection{The roles of $\boldsymbol{<\bar u u + \bar d d>}$ and
$\boldsymbol{<\bar u u - \bar d d>}$}
\label{subsec:condensates}

$<\bar u u + \bar d d>\not=0$ is the signal for what is commonly called
``chiral symmetry breaking'', the breaking of $SU(2)_L
\times SU(2)_R$ down to the diagonal $SU(2)$.
 $<\bar u u-\bar d d> \not=0$ breaks $SU(2)_L$,
 and the custodial $SU(2)$ down to $U(1)_{em}$.
Let us show that $<\bar u u>$ cannot be equal to $<\bar d d>$.
Indeed, for $\nu^3=0$ one gets from 
 (\ref{eq:params}) $\delta_2=0=\kappa_{12}$.
Then 
\begin{equation}
m_{12}^2 = \delta_{12} = \frac{g f_\pi m_\pi^2}{2m_W},
\end{equation}
 in which we used the definition  of $\delta_{12}$ in
(\ref{eq:params}), the GMOR relation (\ref{eq:GMOR}) and (\ref{eq:lambda1}).
 
Performing the minimization of the effective potential $V_{eff}(K,H)$ 
while still
supposing that $V(K,H)$ is $U(2)_L \times U(2)_R$ invariant gives
the two equations
\begin{equation}
m_{H}^2 = \lambda_{H} <{\mathfrak s}^0>^2 +2\delta_1
+\delta_{12}\frac{<{\mathfrak s}^3>}{<{\mathfrak s}^0>},\quad
m_{H}^2 = \lambda_{H}<{\mathfrak s}^3>^2 
+\delta_{12} \frac{<{\mathfrak s}^0>}{<{\mathfrak s}^3>},
\label{eq:minpopt}
\end{equation}
which yield, since $<{\mathfrak s}^3> \gg <{\mathfrak s}^0>$ (see
(\ref{eq:v}) and (\ref{eq:sigma}))
\begin{equation}
\lambda_H \approx \frac{2\delta_1}{<{\mathfrak s}^3>^2} +
\frac{\delta_{12}}{<{\mathfrak s}^0><{\mathfrak s}^3>}
=\frac32\frac{g^2 m_\pi^2}{m_W^2}.
\end{equation}
The mass matrix of the ${\mathfrak s}^0-{\mathfrak s}^3$ system becomes then
(we use (\ref{eq:minpopt}))
\begin{equation}
\left(\begin{array}{cc}
\frac{\partial^2 V_{eff}}{(\partial {\mathfrak s}^0)^2}
\equiv 2\lambda_H <{\mathfrak s}^0>^2-\delta_{12}\frac{<{\mathfrak s}^3>}{<{\mathfrak s}^0>}  &
\frac12\frac{\partial^2 V_{eff}}{\partial {\mathfrak s}^0 \partial
{\mathfrak s}^3}
=0\cr
\frac12\frac{\partial^2 V_{eff}}{\partial {\mathfrak s}^0 \partial
{\mathfrak s}^3}
=0 &
\frac{\partial^2 V_{eff}}{(\partial {\mathfrak s}^3)^2}
\equiv 2\lambda_H <{\mathfrak s}^3>^2-\delta_{12}\frac{<{\mathfrak s}^0>}{<{\mathfrak s}^3>}
\end{array}\right)
\approx
\left(\begin{array}{cc}
- m_\pi^2 & 0   \cr
        0 & 6 m_\pi^2 \end{array}\right).
\end{equation}
It exhibits, because of the term
$ -\delta_{12}\frac{<{\mathfrak s}^3>}{<{\mathfrak s}^0>}$ in
$\frac{\partial^2 V_{eff}}{(\partial {\mathfrak s}^0)^2}$,
which comes from the low energy expression of Yukawa couplings,
  a tachyonic $s$ Higgs boson $m_\varsigma^2 \approx - m_\pi^2$.
The theory with $<\bar u u> = <\bar d d>$ is thus unstable.

Since we have everywhere supposed that the minimum of the effective
potential fits the $\vec W$ and $\vec \pi$ masses, 
we conclude that chiral and weak symmetry breakings as they are observed
are only possible for $<\bar u u> \not= <\bar d d>$.

Unlike for the pions the masses of which are related to $<\bar u u + \bar d
d>$ by the GMOR relation (\ref{eq:GMOR}), there is no such relation between
$m_W$ and $<\bar u u-\bar d d>$ (see the last line of (\ref{eq:sumparams})).
Moreover, even when $<\bar u u> = <\bar d d>$ (that is, $<\nu^3>=0$) 
$<{\mathfrak s}^3>$ can be equal to $\sigma/\sqrt{2}$ because, in its
expression (\ref{eq:compdoub}),
$\nu^3$ cancels between the numerator and the denominator.
This is why it
looks opportune to rather speak of $<\bar u u> \not= <\bar d d>$ as the
{\em catalyst} of weak (and custodial) symmetry breaking.

\subsection{The custodial $\boldsymbol{SU(2)}$}
\label{subsec:custodial}

While $(\bar u u + \bar d d)$ gets annihilated by all generators of $SU(2)$,
$(\bar u u - \bar d d)$ does not, it only gets annihilated by $T^3 = Q$
(see \cite{Machet1}). So,
$<\bar u u> \not= <\bar d d>$ spontaneously breaks the custodial $SU(2)$ down to
$U(1)_{em}$. 
In this breaking one expects two Goldstones. They are the excitations by $T^+$
and $T^-$ of the ${\mathfrak s}^3$ vacuum , that is the two scalars
 ${\mathfrak s}^+$ and ${\mathfrak s}^-$ eaten by $W^\pm$, and which
coincide with the two charged Goldstones of the spontaneously broken weak
$SU(2)_L$.

The electroweak Lagrangian is invariant by the custodial $SU(2)$
 as soon as the $\vec W$'s form an $SU(2)$ vector. But, in the broken phase,
the $W^3$ can only eat ${\mathfrak s}^0$ which is a $SU(2)$
singlet.  This is how the generation of the $\vec W$ mass breaks the custodial
symmetry.

\subsection{Goldstones and pseudo-Goldstones}
\label{subsec:golds}

Three true Goldstones are eaten by the $\vec W$'s
to get massive: they are ${\mathfrak p}^0$, ${\mathfrak s}^+$ and
${\mathfrak s}^-$, belonging to the doublet $H$.
${\mathfrak p}^0$ is also the Goldstone of the $U(1)_L \times U(1)_R$
spontaneous breaking down to the diagonal $U(1)$.
The three $\vec{\mathfrak p}$ (the three pions) are the pseudo-Goldstones
of the broken $SU(2)_L \times SU(2)_R$ down to $SU(2)$.

The only non-Goldstones are the two Higgs bosons $\xi$ and $\varsigma$ 
in the sense that,
though their masses also vanish with $m_\pi$, they do not
seem connected with the breaking of any continuous symmetry.
The first could only be excited by acting either on ${\mathfrak p}^0$ with
$T^3_L$ or $T^3_R$, or on ${\mathfrak p}^3$ with ${\mathbb I}_L$ or
${\mathbb I}_R$. However, in a first approximation, neither ${\mathfrak
p}^0$ nor ${\mathfrak p}^3$, being a pseudoscalar, has a non-vanishing VEV.
Likewise, $H$ could only be excited either by acting on ${\mathfrak p}^3$
with $T^3_L$ or $T^3_R$, or on ${\mathfrak p}^0$ with  ${\mathbb I}_L$ or
${\mathbb I}_R$. The same argumentation rejects thus both as Goldstone
bosons, unless some additional  spontaneously broken continuous symmetry is
at work,  which is to be uncovered.

\section{A few hints for more generations}
\label{section:moregen}

Before concluding, it is worth pointing at  a few features concerning the
case of a larger number $N$ of generations (some information can also be found
in \cite{Machet}). A more detailed study is postponed to \cite{Machet3}.

There are features of this work which only belong to the case of one
generation. For example the fact that the $\eta$ pseudoscalar meson
(pseudoscalar singlet) becomes
the longitudinal neutral $W^3$. In the case of more generations, it may 
happen that this role is still held by the singlet
 $\propto \bar u\gamma_5 u + \bar d\gamma_5 d + \bar
c\gamma_5 c + \bar d\gamma_5 s + \ldots$, but it is no longer the $\eta$,
or by another neutral combination. Though this can only be known by a precise study,
it is likely that the
$\eta$ can then live again its life as a physical  pseudoscalar meson.

Other features are certainly, at the opposite, robust, like the fact that there is a
very light Higgs boson with mass $\leq f_\pi \approx 93\,MeV$. Likewise,
from the expression (\ref{eq:lambdaeff}) for the quartic Higgs
coupling $\lambda_H$, it seems reasonable to  believe that, even if the mass of the
pion gets replaced by the mass of a much heavier bound state, $\lambda_H$
will stay smaller than $1$ and thus ``perturbative''. It can only get equal
to $1$ if $m_\pi$ is replaced by $\sqrt{2}m_W/g \approx 168\,GeV$, such that one
should only be careful when the ``top'' generation is concerned, for which
``non-perturbative'' phenomena could appear.

The logic of the present work and of \cite{Machet1} is that all
(pseudo)scalar doublets isomorphic to the one of the Standard Model of
Glashow, Salam and Weinberg \cite{GSW} should be incorporated. This would stay an empty or
meaningless statement without noticing that the standard Higgs doublet has
transformations by the chiral group (\ref{eq:ruleL}) (\ref{eq:ruleR})
 that are identical to those of bilinear
quark operators. For one generation, this doubled the number of
possible doublets, with parity distinguishing the two of them. 
In the case of $N$ generations, it was shown in \cite{Machet} that there
exists $2N^2$ such doublets, divided, by parity again, in two sets. Their
$8N^2$ real components can be put in one-to-one relationship with the same
number of scalar and pseudoscalar $J=0$ mesons that occur for $2N$ flavors
of quarks. The same logic as the one followed here requires accordingly
that the Standard Model be then endowed with $2N^2$ complex Higgs doublets.
Among these, one expects in particular as many Higgs fields as there exist
quark-antiquark $<\bar q_i q_i>$ condensates,
that is, $2N$.
Owing to the large number of parameters involved, it looks of course too
optimistic to think that one can easily calculate  all masses and
couplings as we did here. This path stays nevertheless in our opinion the most
natural to follow, the underlying guess being that the
mystery of Higgs boson(s) simply lies inside the one of scalar (and eventually
pseudoscalar) $J=0$ mesons.

\section{Conclusion and prospects}
\label{section:conclusion}

As we re-built it, the Standard Model for one generation of fermions
is complete in the sense
that all masses and couplings of all fields present in the Lagrangian and
of all $J=0$ pseudoscalar mesons are determined. Pions are accounted for with the 
correct decays and, of the four expected
scalar mesons, two (the charged ones) become the longitudinal charged
$W^\pm$  while the last two are the Higgs bosons $\varsigma$ and $\xi$. Both have small
masses and are perturbatively coupled and self-coupled. While $\xi$ is
expected to be close to standard, $\varsigma$ is extremely light
and has  peculiar properties that
deserve a specific investigation concerning the role that it can hold in
nature \cite{Machet4}.
As far as we can see, this minimal extension of the Standard Model is
different from what other authors have been considering; it is different as
a 2-Higgs doublet model \cite{BFLRSS} \cite{HHG} \cite{DiazSanchez},
 and it is different in
that, for a larger number of generations $N>1$, it cannot stay as a 2-Higgs
doublet model and should be endowed with $2N^2$ Higgs doublets.
A key ingredient to account
simultaneously for the different scales in presence, weak and chiral, is
parity doubling.
It could  only be uncovered through the one-to-one correspondence
demonstrated in \cite{Machet1} between the Higgs fields and bilinear
quark operators and detailed  symmetry considerations.
The breaking of parity has reflected here in the mass
splitting of the two Higgs bosons, their ratio being precisely that of the two
scales in presence.

At this stage, no physics ``beyond the Standard Model'' looks needed
\footnote{The only hint in favor of it may be the vanishing of the masses of
the two Higgs bosons at the chiral limit, which makes them appear ``like
pseudo-Goldstone bosons'' (see subsection \ref{subsec:golds}).}
 but, since the one generation case can only be considered as a
``toy Standard Model'', this is one among the features that should be
carefully scrutinized for more generations of fermions \cite{Machet3}.

\medskip
{\em \underline{Acknowledgments:} it is a great pleasure to thank
 O.~Babelon, M.~Capdequi-Peyran\`ere, S.~Davidson, M.~Knecht, J.~Lavalle,
 G.~Moultaka, P. Slavich and M.I.~Vysotsky for conversations, advice, and
helping me to correct mistakes.}

\newpage

\begin{em}

\end{em}


\begin{thebibliography}{50}

\bibitem{Machet1}
B.~Machet: ``Unlocking the Standard Model. I. 1 generation of quarks.
 Symmetries'', arXiv:1206.2218 [hep-ph].

\bibitem{GSW}
S.L.~Glashow: "Partial-symmetries of weak interactions". Nuclear
Physics 22 (1961) 579-588;

S.~Weinberg: "A Model of Leptons". Phys. Rev. Lett. 19 (1967): 1264-1266;

A.~Salam: in: Elementary Particle Physics:
 Relativistic Groups and Analyticity; Eighth Nobel Symposium. N. Svartholm.
ed. (Almquvist and Wiksell, Stockholm  1968) pp. 367.

\bibitem{Dashen}
S.L.~Adler \& R.F.~Dashen, ``Current Algebras and Applications to Particle
Physics'', Frontiers in Physics, Benjamin (New York, Amsterdam) 1968.

\bibitem{Lee}
B.W.~Lee: ``Chiral Dynamics'', Gordon Breach, 1972.

\bibitem{dAFFR}
V.~De~Alfaro, S.~Fubini, G.~Furlan \& C.~Rossetti: ``Currents in Hadron
Physics'', North-Holland Publishing Company, 1973.

\bibitem{GMOR}
M.~Gell-Mann, R.~Oakes \& J.~Renner: ``Behavior of Current Divergences
under SU(3) x SU(3)'', Phys.~Rev.~175 (1968) 2195.

\bibitem{Susskind}
L.~Susskind: ``Dynamics of spontaneous symmetry breaking in the
Weinberg-Salam theory'', Phys.~Rev.~D 20 (1979) 2619.

\bibitem{Machet}
B.~Machet: ``Chiral scalar fields, custodial symmetry in electroweak
$SU(2)_L\times U(1)$ and the quantization of the electric charge'', Phys.
Lett. B 385 (1996) 198-208.

\bibitem{Machet3}
B.~Machet: ``Unlocking the Standard Model. III. The case of $N>1$
generations'', in preparation.

\bibitem{Machet4}
B.~Machet: ``Unlocking the Standard Model. IV. A
natural candidate for dark matter'', in preparation.

\bibitem{BFLRSS}
See for example in:\newline
G.C.~Branco, P.M.~Ferreira, L.~Lavoura, M.N.~Rebelo, M.~Sher \& J.P.~Silva:
``Theory and phenomenology of two-Higgs-doublet models'', Physics
Reports (2012), in press, and references therein

\bibitem{HHG}
J.F.~Gunion, H.~Haber, G.~Kane \& S.~Dawson: ``The Higgs
Hunter's Guide'' Westview Press, Perseus Books (2000).

\bibitem{DiazSanchez}
R.A.~Diaz~Sanchez: ``Phenomenological analysis of the two Higgs doublet model'',
Ph. D. thesis at Univ. National of Colombia (Bogota), hep-ph/0212237 (2002).


\end{thebibliography}
\end{document}